\documentclass[journal=aesccq,manuscript=article,layout=twocolumn]{achemso} %twocolumn or the default layout is traditional
\usepackage[fontsize=11pt]{fontsize}
\setkeys{acs}{doi = true}

\usepackage{chemformula} % Formula subscripts using \ch{}
\usepackage[T1]{fontenc} % Use modern font encodings

\usepackage[pdftex,plainpages=false,
    pdfpagelabels,
    pdfusetitle,
    bookmarksnumbered,
    linkbordercolor={1 1 1},
    colorlinks={ true },
]{hyperref}
\usepackage[version=4]{mhchem}
\usepackage{tabularx}
\usepackage{booktabs}   %for \toprule
\usepackage{chemfig}

\newcommand{\doi}[1]{\href{http://dx.doi.org/#1}{\nolinkurl{#1}}}

\newcommand{\eV}{{\ensuremath{\textrm{eV}}}}

\newcommand{\massU}{{\ensuremath{m/z}}}

\newcommand{\rateU}{{\ensuremath{\text{cm}^{3}\,\text{s}^{-1}}}}

\newcommand{\IE}{{\ensuremath{\textrm{IE}}}}

\newcommand{\HH}{{\ce{H2}}}
\newcommand{\DD}{{\ce{D2}}}

\newcommand{\ie}{{i.\,e.}}

\newcommand{\Clp}{\ce{Cl+}}
\newcommand{\HHClp}{\ce{H2Cl+}}
\newcommand{\HClp}{\ce{HCl+}}

\newcommand{\HP}{\hphantom}

\newcommand{\cHHblue}{\chemfig[atom sep=1.8em]{\color{blue}{H}-\color{blue}{H}}}
\newcommand{\cHClred}{\chemfig[atom sep=1.8em]{\color{red}{H}-\charge{[extra sep=0pt]45[anchor=180+\chargeangle]=$\scriptstyle+$}{Cl}}}
\newcommand{\cHHClredblue}{\chemfig[atom sep=1.8em]{\color{red}{H}-[:45.4]\charge{[extra sep=0pt]45
     [anchor=180+\chargeangle]=$\scriptstyle+$}{Cl}-[:-45.4]\color{blue}{H}}}
\newcommand{\cHHClblueblue}{\chemfig[atom sep=1.8em]{\color{blue}{H}-[:45.4]\charge{[extra sep=0pt]45
     [anchor=180+\chargeangle]=$\scriptstyle+$}{Cl}-[:-45.4]\color{blue}{H}}}

\author{Miguel Jim\'enez-Redondo}
\affiliation[mpe]
{Max Planck Institute for Extraterrestrial Physics, Giessenbachstrasse 1, 85748 Garching, Germany}
\author{Olli Sipil\"a}
\affiliation[mpe]
{Max Planck Institute for Extraterrestrial Physics, Giessenbachstrasse 1, 85748 Garching, Germany}
\author{Robin Dahl}
\affiliation[mpe]
{Max Planck Institute for Extraterrestrial Physics, Giessenbachstrasse 1, 85748 Garching, Germany}
\alsoaffiliation[bonn]
{Mulliken Center for Theoretical Chemistry, University of Bonn, Beringstrasse 4, 53115 Bonn, Germany}
\alsoaffiliation[rwth]
{Institute of Physical Chemistry, RWTH Aachen University, Landoltweg 2, 52074 Aachen, Germany}
\author{Paola Caselli}
\affiliation[mpe]
{Max Planck Institute for Extraterrestrial Physics, Giessenbachstrasse 1, 85748 Garching, Germany}
\author{Pavol Jusko}
\email{pjusko@mpe.mpg.de}
\affiliation[mpe]
{Max Planck Institute for Extraterrestrial Physics, Giessenbachstrasse 1, 85748 Garching, Germany}

\title[Cl+ and HCl+]
  {\ce{Cl+} and \ce{HCl+} in Reaction with \ce{H2} and Isotopologues: a Glance into H-abstraction and Indirect Exchange at Astrophysical Conditions}% \footnote{A footnote for the title}}


\begin{document}

\begin{abstract}
Astrochemical models of interstellar clouds, the sites of stars and planet formation,
require information about spin-state chemistry to allow quantitative comparison with
spectroscopic observations. In particular, it is important to know if full scrambling
or H-abstraction (also known as proton hop) takes place in ion-neutral reactions.
The reaction of \ce{Cl+} and \ce{HCl+} with \ce{H2} and isotopologues has been studied at cryogenic
temperatures between $20-180$~K using a 22 pole radio-frequency ion trap.
Isotopic exchange processes are used to probe the reaction mechanism of
the \ce{HCl+ + H2} reaction. The results are compared to previous measurements and theoretical predictions.
The rate coefficients
for the \ce{Cl+ + H2} and \ce{HCl+ + H2} reactions are found to be constant in the range of temperatures studied,
except for the \ce{DCl+ + D2} reaction, where a weak negative temperature dependence is observed, and
reactions with \ce{D2} are found to be significantly slower than the Langevin rate. No isotopic exchange reactions are
observed to occur for the \HHClp\ ion. The analysis of the products of the \ce{HCl+ + H2} isotopic system
clearly indicates that the reaction proceeds via a simple hydrogen atom abstraction.
\end{abstract}

\section{Introduction}
Despite its low elemental abundance, chlorine chemistry presents some peculiarities that make it an interesting
target for astrochemical studies. The chlorine atom can easily be ionised by UV radiation
\begin{equation}
 \ce{Cl} + h\nu \ch{->} \ce{Cl+} + \ce{e^-},   ~~~~~
 \IE=12.97\;\eV      \label{eq:photo}
\end{equation}
and has an ionization energy (\IE) that is below the ionization energy
of the \ce{H} atom
($\IE(\ce{H})=13.60\;\eV$). As a consequence, \ce{Cl+} does not
charge transfer to the most abundant atom in space, the \ce{H} atom, and therefore chlorine is predominantly found in its ionized form in the diffuse interstellar medium. Of all elements with this behavior, chlorine is unique in that the singly charged cation, \Clp, reacts exothermically with \HH\ \cite{Neufeld2009}
\begin{equation}
    \ce{Cl+} + \ce{H2} \ch{->} \HClp + \ce{H} \quad \Delta H = -0.175\;\eV \cite{GlenewinkelMeyer1991},         \label{eq:HCl}
\end{equation}
producing a very reactive \HClp\ ion, which quickly undergoes another reaction with \HH\
\begin{equation}
    \ce{HCl+} + \ce{H2} \ch{->} \HHClp + \ce{H} \quad \Delta H = -0.423\;\eV \cite{LeGal2017},         \label{eq:HHCl}
\end{equation}
forming a rather non reactive closed shell molecular ion, \HHClp, that does not react further with \HH.

The H$_2$Cl$^+$ ion was first detected using the Herschel Heterodyne Instrument for the Far-Infrared (HIFI) towards NGC 6334I and Sgr B2(S)\cite{Lis2010}.
It has also been observed in the Orion Bar, Orion S, W31C, Sgr A\cite{Neufeld2012}, the massive star-forming regions W31C and W49N\cite{Gerin2013}, and extragalactic sources using the Atacama Large Millimeter/submillimeter Array (ALMA) \cite{Muller2014,LeGal2017}.
Typical column densities for this ion are in the range of $N(\ce{H2Cl+})/N(\ce{H}) = 0.9$--$4.8 \times 10^{-9}$ \cite{Muller2014}.
Herschel observations towards several sources (G29.96-0.02, W49N, W51, W3(OH)) allowed the determination
of the ortho-para and \ce{^{35}Cl/^{37}Cl} isotopic ratios \cite{Neufeld2015}.
\ce{HCl+} was first observed using Herschel/HIFI towards
W31C (G10.6-0.4) and W49N \cite{DeLuca2012},
and in W49N using the GREAT instrument on board SOFIA \cite{Neufeld2021},
with column densities with respect to atomic hydrogen of $2$--$9 \times 10^{-9}$
\cite{DeLuca2012,Neufeld2021}. The \ce{Cl+} ion has been mainly observed in diffuse clouds,
with a wide range of column densities, from $10^{-9}$ to $10^{-7}$ with respect to atomic hydrogen
\cite{Moomey2011,Ritchey2023}.
%\textcolor{blue}{Add this reference from Paola here \cite{Ritchey2023}.}
Despite the relative simplicity of the
chlorine hydrides chemistry, astrochemical models have issues reproducing the observed abundances of HCl,
\ce{HCl+} and \ce{H2Cl+} \cite{Neufeld2009,Acharyya2017,Neufeld2021}, which is usually attributed to the lack
of an accurate dissociative recombination rate coefficient for \ce{H2Cl+}.

Reaction (\ref{eq:HHCl}) controls the ortho-para ratio of the \HHClp\ ion, which is greatly affected by the
specific reaction mechanism. The reaction of \HClp\ with \HH\ can in principle proceed via two
different mechanisms, a direct H-abstraction
\begin{equation}
\schemestart
\cHClred \+{8pt,4pt,1pt} \cHHblue \arrow{}[,0.7] \cHHClredblue \arrow{0}[,0] \+{8pt,4pt,0pt} \color{blue}{H}
\schemestop
\label{eq:hop}
\end{equation}
or full scrambling (indirect exchange)
\begin{equation}
\schemestart
\cHClred \+{8pt,4pt,1pt} \cHHblue \arrow{}[,0.7] \cHHClblueblue \arrow{0}[,0] \+{8pt,4pt,0pt} \color{red}{H}
\schemestop
\label{eq:scrambling}
\end{equation}
This allowed \citet{LeGal2017} to infer the dominant reaction mechanism, H-abstraction, based on the
astronomical observations of the ortho-para ratio of \HHClp, complementing it with a theoretical study supporting the
observational results. The evidence of this reaction proceeding through the direct mechanism
contrasts with the well studied \ce{H3+ + H2} system, where the nuclear spin dependence of
the reaction has been studied from both the experimental and theoretical points of view
\cite{Cordonnier2000,Crabtree2011,Suleimanov2018} showing a preference for the full scrambling mechanism at
low temperatures. Data on the specific reaction mechanism is limited for most reactions of interest in
astrochemistry, while having a significant impact on models considering spin-state chemistry and/or
deuteration \cite{Sipila2019}.

The \HClp\ and \HHClp\ ions have been characterized in the laboratory in a variety of ways.
Rotational transitions of \HClp\ has been studied by far-infrared laser magnetic resonance
\cite{Lubic1989} and later in high resolution in an microwave discharge absorption
experiment \cite{Gupta2012} leading to its first
detection in space \cite{DeLuca2012}. Correspondingly, the detection of \HHClp\ \cite{Lis2010} was based
on the submillimeter-wave spectra of \HHClp\ recorded by \citet{Araki2001} in a hollow-cathode discharge.
The dissociative recombination rate coefficient of \HHClp\ with electrons
has been determined in a pulsed discharge
\cite{Kawaguchi2016}, while cryogenic storage ring experiments have been used to determine the corresponding
rate coefficients for \HClp\ \cite{Novotny2013} and \ce{D2Cl+} \cite{Novotny2018}.

The reactivity of the \HClp\ and \HHClp\ ions with \HH\
has been extensively studied using flowing-afterglow (FA)\cite{Fehsenfeld1974},
selective-ion flow tube (SIFT)\cite{Raouf1980,Rakshit1980,Smith1981,Hamdan1982,Mayhew1990}, and
ion-cyclotron resonance (ICR)\cite{Cates1981, Kemper1983}.
In all these works, the reaction \HClp\ + \HH\ is faster than \Clp\ + \HH, but there is a large spread
in the measured rate coefficients, especially for the \ce{HCl+ + H2} reaction,
with more than a factor of 2 difference between the lowest and highest values.
Three of the studies
examined the temperature dependence of the rate coefficients down to 80~K \cite{Smith1981}
and 150~K \cite{Cates1981,Kemper1983}. For the \ce{Cl+ + H2} reaction, \citet{Cates1981} reported a positive
temperature dependence, later reexamined by \citet{Kemper1983} in the same setup and finding a weak negative temperature dependence in line with the results of \citet{Smith1981}. For the \ce{HCl+ + H2} reaction, the ICR
measurements show a weak negative temperature dependence \cite{Cates1981,Kemper1983}, while the SIFT
experiment shows no temperature dependence \cite{Smith1981}. No
measurements of the rate coefficients below 80~K have been reported previously.

Limited theoretical
studies are available for these systems. The chemiluminescence of the \ce{Cl+ + H2} reaction
has been studied using a beam-target gas collision setup \cite{GlenewinkelMeyer1991},
where the authors also present theoretical calculations for the metastable \Clp\ ion
\cite{GlenewinkelMeyer1991a}. The reaction of \HClp\ with \HH\ has been studied theoretically by \citet{LeGal2017}, focusing on the prevalence of H-abstraction
with respect to full scrambling.

In this work, we present a study of the \ce{Cl+ + H2} and \ce{HCl+ + H2} isotopic systems at cryogenic temperatures using a 22 pole radio-frequency ion trap. The rate coefficients of the reactions are determined in the temperature range of $20-180\;\text{K}$. Isotopic exchange processes are used to probe the reaction mechanism of the \ce{HCl+ + H2} reaction. The results are compared to previous measurements and theoretical predictions.

\section{Experimental Section}

\begin{figure}[h]
\centering
  \includegraphics[width=0.99\linewidth]{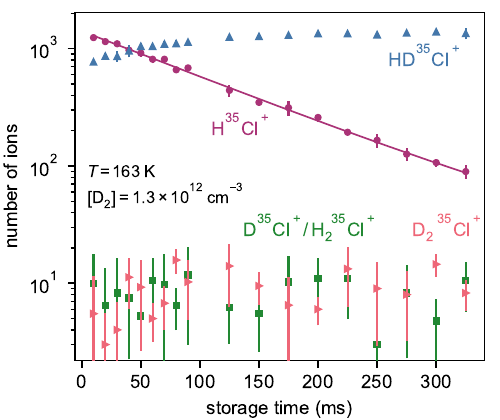}
  \caption{Time evolution of the number of ions in the trap for the reaction of
  \HClp\ with \DD. Purple line represents the fit of the \HClp\ signal to an
exponential decay. Storage time ``0'' represents the time at which the trap
is closed after filling. Some ions react during the trap filling stage, leading to the
non-zero signal of \ce{HDCl+} at 0 ms.} \label{f_one}
\end{figure}

All the experiments were conducted in a cryogenic 22 pole radio-frequency ion trap setup, CCIT. We provide only a short description here,
as the setup has been extensively described elsewhere \cite{Jusko2024}.
No unexpected or unusually high safety hazards were encountered.
The ions \ch{Cl+}, \ch{HCl+}, and \ch{DCl+}, were produced in a storage ion source \cite{Gerlich1992} using electron bombardment
of hydrochloric acid (water solution $37\,\%$, Carl Roth GmbH) vapors introduced into the setup through a variable leak valve.
\ch{DCl} in \ce{D2O} has been used in case of \ch{DCl+} ions.
Electron impact ionization of \ch{HCl} \cite{Harper2001} produces all the possible ions,
that is, \ch{Cl+}, \ch{HCl+}, and \ch{H2Cl+} for both isotopes of \ch{^{35}Cl} and \ch{^{37}Cl}.
Therefore, the ion of the mass-charge  ratio ($\massU$) of interest is selected in a
quadrupole mass filter, prior to the injection into the 22 pole trap, where intense \ch{He} pulse is used to store and cool it down
close to the temperature of the trap walls.
The reactant gas, either \ch{H2} or \ch{D2} (Westfalen, no further purification), is leaked
into the trap continuously.
Normal, \ie, room temperature \HH\ (3:1 ortho:para) and \DD\ (2:1 ortho:para)
are used in all the experiments (note that lower energy spin isomer of \ce{H2} is para, but it is ortho for \ce{D2}).
The pressure is recorded using a Bayard-Alpert ion gauge,
which is calibrated against an absolute pressure gauge (membrane baratron, CMR 375, Pfeiffer), and, at each
given temperature, the corresponding number density is determined (see \citet{Jusko2024} for details).
After a variable storage time, that is, variable exposure to the neutral gas, the primary and product ions are extracted from the
trap towards a second quadrupole and counted in a Daly type detector. The measurement with the shortest storage time is performed a few ms after the trap is closed, to ensure the He buffer gas is sufficiently evacuated from the trap.
Only one mass can be filtered at the time, therefore, the experiment is repeated for every primary and product ion mass until
sufficient statistical signal to noise is achieved (see Fig.~\ref{f_one} for one particular temperature and neutral gas number density).
The rate $r$ (inverse of the lifetime $\tau$) of reaction of the primary ion with the neutral gas is determined using
an exponential decay least-square fit of the primary ion (full line in Fig.~\ref{f_one}).
This procedure is repeated at several neutral reactant gas pressures, that is,
number densities $\left[ X \right]$ ($\left[ \ch{H2} \right]$ or $\left[ \ch{D2} \right]$).
The final reaction rate coefficient $k$ is then determined as a parameter of the linear least square fit in the form of
\begin{equation}
    r = k \left[ X \right]+C,
\end{equation}
which produces more reliable results than using a single pressure point, and can account
for spurious ion losses through the offset $C$ (see refs.~\cite{Jusko2024,Dohnal2023}
for further details).
The uncertainties in the rate coefficient values reported throughout the text correspond to the
statistical error of this fit. The main source of uncertainty, however,
is the systematic error introduced
by the pressure calibration, which is estimated to be 20\% \cite{Jusko2024}. This
process is repeated for every ion-neutral pair of interest at the different temperatures
sampled in the $20-180$~K range.

For the isotopic exchange reactions, that is, reactions of the \ce{H2Cl+ + H2} isotopic system, the (lack of)
reactivity was probed by introducing a large amount of the corresponding neutral reactant gas into the trap.
The ionic species were produced in different ways depending on the particular reaction. For
\ce{HDCl+ + H2 / D2}, the ions formed during the study of the \ce{DCl+ + H2} and \ce{HCl+ + D2} rate coefficients,
respectively, were used. For \ce{D2Cl+ + H2}, \ce{D2{}^37Cl+}
was directly produced in the ion source, as there is no mass overlap with the other
\ce{H_xCl+} isotopologues.
Finally, \ce{H2Cl+ + D2} was probed during the study of the \ce{DCl+ + D2} reaction, as there is
a 10\% fraction of the ions at mass 37 that does not react with \ce{D2}, which corresponds to \ce{H2Cl+} ions
produced in the ion source from residual \ce{H2}.
These reaction rate coefficients are reported as upper limits, that is, the highest value of the rate
coefficient within the uncertainty of the fit is displayed in Table~\ref{tab:kida},
as no reaction was observed within the experimental time frame.

\section{Results and Discussion}

\begin{figure}[h]
\centering
  \includegraphics[width=0.99\linewidth]{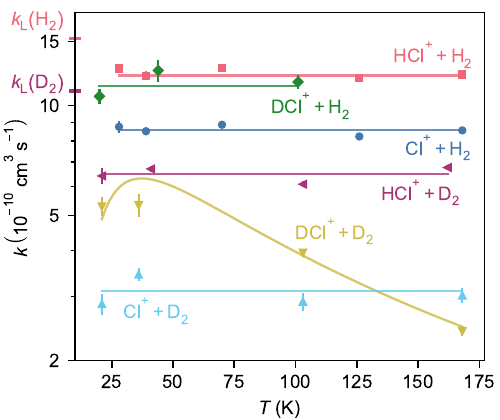}
  \caption{Temperature dependence of all the rate coefficients determined in this
  work. Marks on the vertical axis represent the values of the Langevin rate
  coefficient $k_L$ for reactions with \HH\ or \DD.}\label{f_all}
\end{figure}

The full set of rate coefficients measured in this work is shown in Fig.~\ref{f_all}. The rate coefficients
of all reactions are found to be constant in the range of temperatures studied ($20-180\,\text{K}$), except in
the case of the \ce{DCl+ + D2} reaction, where a weak negative temperature dependence is observed. The values
of the rate coefficients for the \ce{HCl+ + H2} and \ce{DCl+ + H2} reactions are close to the
corresponding Langevin rate coefficient ($k_L(\ce{H2}) = 1.53 \times 10^{-9}$~\rateU),
while most other reactions,
especially those involving \ce{D2} ($k_L(\ce{D2}) = 1.10 \times 10^{-9}$~\rateU),
are significantly slower than their Langevin counterpart.

\begin{table*}[htb]
    \caption[]{%Parameterized
    Experimentally determined reaction rate coefficients of reactions (\ref{eq:HCl}) and (\ref{eq:HHCl})
    from Fig.\,\ref{f_all} between $20-170\;\text{K}$ for reactions with \ce{H2} and $20-280\;\text{K}$ for reactions with \ce{D2}.
    }\label{tab:kida}
    \begin{center}
    \begin{tabular}{llr@{\HP}lcl} % S[table-format = 3.2(3)] S[table-format = 4.3(3)] S[table-format = 2.0(2)] c c}
       \toprule
       \multicolumn{2}{c}{Reaction}  & \multicolumn{1}{c}{$k^a$} & & Previous works$^{a,b}$ \\
       \midrule
       \ch{Cl+ + H2} & \ch{-> HCl+ + H}    & $8.56  \pm 0.12$      &      & $6.5 \pm 1.0$\cite{Kemper1983}, $8.7 \pm 1.7\cite{Smith1981}$, $7.1 \pm 2.1\cite{Fehsenfeld1974}$, $9.2 \pm 1.8\cite{Raouf1980}$                             \\
        &  &    &      & $5.5 \pm 1.0\cite{Cates1981}$, $8.3 \pm 1.6\cite{Mayhew1990}$, $6.8 \pm 2.0\cite{Rakshit1980}$                                 \\
\ch{Cl+ + D2}  & \ch{-> DCl+ + D}    & $3.09  \pm 0.14$      &      & $4.5 \pm 0.5\cite{Kemper1983}$, $6.0 \pm 1.2\cite{Smith1981}$ (80~K), $4.9 \pm 1.0\cite{Mayhew1990}$                                    \\
\ch{HCl+ + H2} & \ch{-> H2Cl+ + H}   & $12.10 \pm 0.13$      &      & $8.6 \pm 0.8\cite{Kemper1983}$, $13 \pm 2.6\cite{Smith1981}$, $5.2 \pm 1.6\cite{Fehsenfeld1974}$, $5.5 \pm 1.0\cite{Cates1981}$, $8.9 \pm 1.8\cite{Hamdan1982}$                                   \\
\ch{HCl+ + D2} & \ch{-> HDCl+ + D}   & $6.50  \pm 0.18$      &      &                                    \\
\ch{DCl+ + H2} & \ch{-> HDCl+ + H}   & $11.30 \pm 0.48$      &      &    \\
\ch{DCl+ + D2} & \ch{-> D2Cl+ + D}   & $5.29  \pm 0.32$      & $^c$ &   \\
\ch{HDCl+ + H2}& \ch{-> no~reaction}  & $<6 \times 10^{-4}$  &      & \\
\ch{HDCl+ + D2}& \ch{-> no~reaction}  & $<4 \times 10^{-5}$  &      & \\
\ch{D2Cl+ + H2}& \ch{-> no~reaction}  & $<2 \times 10^{-4}$  &      & \\
\ch{H2Cl+ + D2}& \ch{-> no~reaction}  & $<3 \times 10^{-4}$  &      & \\
      \bottomrule
    \end{tabular}
    \end{center}
    \footnotesize{Note: $a$ -- in $10^{-10}\;\rateU$;
        $b$ -- at $300\;\text{K}$, unless stated otherwise;
    $c$ -- at $20\;\text{K}$, 
    best fit using Arrhenius–Kooij formula \citep{Wakelam2012} $k(T) = A(T/300)^B \exp(-C/T)$:
    $A=1.56\pm0.39$~$\times 10^{-10}\;\rateU$, $B=-1.29\pm0.51$, $C=48\pm21$~K.
    The uncertainties reported correspond to the statistical errors of the least square fit.
    Total uncertainty is estimated to be 20\% (see text).}
\end{table*}

A single product was observed for the mixed isotopic systems, as in the case of \ce{HCl+ + D2} depicted
in Fig.~\ref{f_one}. Only residual counts are detected for mass $39\;\massU$,
which corresponds to \ce{D2Cl+} ions.
This clearly indicates that no scrambling takes place in this system and the reaction proceeds via a simple
deuterium atom abstraction. This very clear experimental evidence is aided by the fact that the
potential subsequent isotopic exchange reactions (\ch{HDCl+ + D2 -> D2Cl+ + HD}
in this particular case) do not occur at a detectable rate in this system, as this
would otherwise lead to the presence of \ch{D2Cl+} in the trap even in the absence of scrambling in the
initial reaction. Our result is in line with the theoretical work of \citet{LeGal2017},
who found that H-exchange processes accounted for less than 1\% of the reactive collisions in
\ch{HCl+ + H2 -> H2Cl+ + H}. The reaction was instead found to proceed overwhelmingly via H-abstraction.
Furthermore, the authors report a high barrier of 0.631~eV for the formation of the \ce{H3Cl+} complex
from \ce{H2Cl+} and H, which is likely responsible for the lack of isotopic exchange observed in our experiment.

\begin{figure}[h]
\centering
  \includegraphics[width=0.99\linewidth]{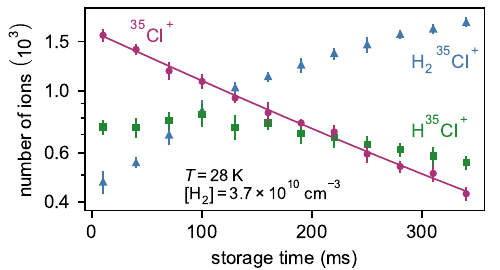}
  \caption{Number of ions in the trap as a function of storage time for the
  reaction of \Clp\ with \HH, producing \HClp, and \HHClp\ in a subsequent reaction.
  A single exponential decay is observed for the reaction of \Clp\ with \HH.}\label{f_Cl_p}
\end{figure}

The rate coefficients determined in this work are summarized in Table~\ref{tab:kida}. Previous available
measurements from FA \cite{Fehsenfeld1974}, SIFT \cite{Raouf1980,Rakshit1980,Smith1981,Hamdan1982,Mayhew1990}
and ICR \cite{Cates1981,Kemper1983}
experiments are also listed for comparison. Most of the measurements in the literature correspond
to the \ce{Cl+ + H2} and \ce{HCl+ + H2} reactions. The rate coefficients for \ce{Cl+ + H2} measured in this work
fall into the higher range of the previously reported values, in good agreement with the SIFT experiment of
\citet{Smith1981} at 300~K. The authors observed a weak negative
temperature trend, measuring a $\sim 15\%$ higher value at 80~K compared to 300~K.
The negative temperature dependence, although with lower rate coefficient values,
was also observed in the ICR experiment of \citet{Kemper1983} between
100--400~K. As mentioned in the introduction, earlier results from the same instrument
\cite{Cates1981} showed a positive temperature dependence. Our measurements show a clear lack of
temperature dependence in the range of $20-180\,\text{K}$ for the reaction of \Clp\ with \HH. A similar
situation occurs for the \ce{Cl+ + D2} reaction, where the measurement by \citet{Kemper1983}
also showed a negative temperature dependence contrasting with our constant rate coefficient. In this case, the
value determined in this work is $\sim$ 30--50\;\% lower than the ones reported previously
\cite{Kemper1983,Smith1981,Mayhew1990}.
Some authors report the presence of metastable (electronically excited) \ce{Cl+(^1D,^1S)} ions
on top of ground state \ce{Cl+(^3P)}
in their experiment under certain conditions \cite{Rakshit1980,Smith1981,GlenewinkelMeyer1991}.
Up to $25\;\%$ of all ions were reported to be in the \ce{Cl+(^1D,^1S)} states and the authors
also claim that metastable \ch{Cl+} is not effectively quenched by the \ch{He} gas \cite{Rakshit1980}.
The theoretically predicted radiative lifetime of these states
is $\sim 10$~s for \ce{Cl+(^1D)} and $\sim 0.5$~s for \ce{Cl+(^1S)},
which decays to \ce{Cl+(^1D)} \cite{Mendoza1983,Biemont1986}.
Moreover, evidence of metastable \ce{O+(^2D,^2P)} ions, with fractions of up to $10\;\%$,
has been shown recently in an experiment using a similar ion source and trap
to ours \cite{Kovalenko2018}.
Although we did not explicitly try to use chemical probing of \ce{Cl+(^1D,^1S)}
with \ch{CO} or \ch{CO2} as shown previously \cite{Rakshit1980}, we are confident
that the presence of metastable \ce{Cl+(^1D,^1S)} is low in our experiment,
as can be seen from Fig.~\ref{f_Cl_p}, where the primary ion \ch{^{35}Cl+}
reacts with \ch{H2} in a single exponential decay.
In case metastable and stable \ch{Cl+} were present in non negligible amounts,
and both reacted with \ch{H2} with a different reaction rate, the
number of ions in the trap should exhibit a double exponential behaviour.

Similarly, the excited electronic state of the \ce{HCl+(^2\Sigma)} is cca. $3.5\;\text{eV}$
higher than its electronic ground state \ce{HCl+(^2\Pi)}.
We do not expect \ce{HCl+(^2\Sigma)} to be present in our experiments, as 
even if the ions were formed in this state in our ion source, the 
transition to the electronic ground 
state is allowed and the ions would radiatively decay
already prior or during the trap filling process\cite{GlenewinkelMeyer1991}.
The situation with \HClp\ is even more complicated, as the \ce{^2\Pi_\Omega} state has
two separate rotational ladders with $\Omega=3/2$ (ground state) and 
$\Omega=1/2$ ($926\;\text{K}$/$0.08\;\text{eV}$ higher)\cite{Gupta2012}.
We expect predominantly the \ce{HCl+(^2\Pi_{3/2})} rotational ladder
to be present in our experiments, 
however, we can not rule out, nor have we tried to experimentally confirm, that even after 
collisional thermalization with the intense He buffer gas pulse at the trap injection, a meaningful 
abundance of \ce{HCl+(^2\Pi_{1/2})} is present. 
At the same time, as in the case of \ce{Cl+}, a significant presence of both states with
noticeably different reactivity would lead to a double exponential behavior that is
not observed experimentally. 

The observed isotope effect on the reaction (\ref{eq:HHCl}), mainly the temperature
dependence of the reaction rate coefficient, is not simply explainable:
1.) the non-deuterated reaction (\ref{eq:HHCl}) is exothermic with only one negative barrier
after the first transition state (see Fig.~4. in \citet{LeGal2017});
2.) the total electronic spin is conserved: \ce{HCl+(^2\Pi) + H2(^1\Sigma) \ch{->} H2Cl+(^1A_1) + H(^2S)};
and; 3.) the difference between the ortho-para (o:p) energy levels of \ce{H2}(3:1) and
\ce{D2}(2:1) in our experiments, where \ce{D2} has a
higher population of the lower energy ortho isomer
and where \ce{H2} has a higher population of the higher energy ortho isomer
(i.e., beneficial for the non-deuterated reaction (\ref{eq:HHCl})),
is negligible in comparison to the other energies at play.
Ultimately, we think a full quantum calculation, as performed for the non-deuterated
reaction (\ref{eq:HHCl}) \cite{LeGal2017} and perhaps experiments with the \ce{HD} isotopologue,
need to be performed to shed light on this isotopic effect.

Finally, the rate coefficient for the \ce{HCl+ + H2}
reaction is again on the higher end of the previously reported values,
in good agreement with the SIFT experiment of \citet{Smith1981},
who also did not observe any temperature dependence between $80-470\;\text{K}$,
contrary to the ICR experiments \cite{Cates1981,Kemper1983}.

\section{Astrochemical implications}

Many deuterium chemical networks used in present-day chemical models include thousands of reactions
(e.g., \citep{Majumdar2017,Hily-Blant2018,Sipila2019}). Such large-scale networks cannot be generated by hand.
Instead, they are typically created by modifying a base chemical network that does not include
fractionation chemistry, and then introducing isotopic substitution via an algorithm.
This automated procedure is based on a set of pre-determined rules, and the most fundamental of these
is the assumption on how the main deuteration reactions actually proceed, that is,
via H-abstraction or full scrambling.
The reaction mechanism affects the number of product branches and the branching ratios
(see for example Fig.\,2 in \citep{Sipila2019}), and hence the predictions of simulations depend on which mechanism
is chosen.
Although the \ch{H3+ + H2} reacting system appears to follow scrambling at low temperatures \citep{Hugo2009},
it is not at all clear that such behaviour will apply universally, and chemical models require constraints from
observations and experiments (such as the present one) to be able to produce reliable predictions.

The full scrambling scenario predicts non-thermal spin-state ratios in the gas phase at low temperatures and
at volume densities appropriate to molecular clouds and dense cores (though approximately statistical ratios
may arise at the late stages of the collapse of cores \citep{Hily-Blant2018}). Reactions on grain surfaces
however mostly proceed via hydrogen/deuterium addition, which leads to statistical spin-state ratios.
Observations of the spin-state ratios of deuterated ammonia \citep{Lis2006,Daniel2016,Harju2017,Wienen2021}
and of \ch{H2Cl+} by \citet{LeGal2017} indicate statistical values.
Therefore, if gas-phase reactions were indeed to proceed via scrambling, the observed statistical ratios
could be explained if a substantial amount of ice is desorbed over the typical lifetime of a cloud.
This scenario appears unlikely especially for ammonia given its high binding energy
on grains \cite{Kakkenpara2024}.
Indeed, the recent observational and simulation study of \citet{Harju2024} shows that the proton hop model
can satisfy the observed values without the requirement of excess desorption, although some still-unexplained
discrepancies between the observed and modelled D/H and spin ratios remain
(this may be due to other factors such as the physical model employed).

The experiments presented here -- together with the theoretical calculations of \citet{LeGal2017} --
provide direct evidence for the proton hop mechanism as the dominant one governing the D/H and spin-state ratios of \ce{H2Cl+}. A similar conclusion was reached recently in the case of ammonia \citet{Harju2024}. Additional experimental and theoretical studies of other reacting systems must be carried out to ascertain whether the dominant role of the proton hop mechanism can be generalized to the formation of other molecules. The \ch{H3+ + H2} reacting system still needs to be treated as a special case, and for that system
the rate coefficients presented by \citet{Hugo2009} represent the state-of-the-art, although most of
their rate coefficients have not been verified experimentally and indeed there is new experimental evidence
suggesting that especially at higher temperatures the rate coefficients are still not known sufficiently well \citep{Jimenez-Redondo2024}.

\section{Conclusions}
The rate coefficients for the isotopic systems \ce{Cl+ + H2} and \ce{HCl+ + H2} have been measured in
a 22 pole radio-frequency ion trap in the temperature range of $20-180$~K. The values obtained
are below the corresponding Langevin rate, particularly for reactions with \ce{D2}, and are
comparable to those available in the literature for higher temperatures, but contrastingly
show no variation with temperature in the range studied. Only the \ce{DCl+ + D2} reaction, for which
no previous data is available, shows a weak negative temperature dependence. The exchange
reactions for the \HHClp\ ion (isotopic variations of \ce{H2Cl+ + H2}) have not been observed to take place
at detectable rates. The product distribution of the isotopic \ce{HCl+ + H2} system clearly indicates that
the reaction proceeds via a simple H-abstraction mechanism, in agreement with the theoretical work of
\citet{LeGal2017}, with no evidence of scrambling.

The results of this work have implications for chemical models of the interstellar medium. The spin-state
ratios and deuteration of molecules are strongly affected by the reaction mechanism, and the present results
show that chlorine hydride chemistry is likely dominated by direct H-abstraction, in contrast to the
\ch{H3+ + H2} system, which is known to proceed via full scrambling at low temperatures. Our results
together with recent observational and modelling efforts suggest that the direct proton hop or
abstraction mechanism is the dominant one at least in the cases of ammonia and \ce{H2Cl+} formation. Whether this result can be generalized to other reacting systems must be verified via additional experimental and theoretical work; the \ch{H3+ + H2} system should always be treated as a special case. Nevertheless, the results of this work suggest that similar ion trap experiments can be used to provide constraints on the reaction mechanisms of other
systems of interest in astrochemistry, extending the limited data available on the topic.

\subsection*{Conflicts of interest}
The authors declare no competing financial interest.

\begin{acknowledgement}

This work was supported by the Max Planck Society.
The authors gratefully acknowledge the work of the electrical and mechanical workshops and engineering
departments of the Max Planck Institute for Extraterrestrial Physics.
We thank Dr. Octavio Roncero for helpful discussions.
We thank the anonymous reviewers for comments enhancing this work.
\end{acknowledgement}

\subsection*{Data availability statement}
The data that support the findings of this study are openly available at {\small \url{https://doi.org/10.5281/zenodo.13741732}}.
\bibliography{main}

\end{document}